\documentstyle[prb,eqsecnum,aps]{revtex}
\twocolumn
\begin{document}
\draft
\twocolumn[
\hsize\textwidth\columnwidth\hsize\csname@twocolumnfalse\endcsname
\title{Proximity effect, quasiparticle transport, and local magnetic moment 
in  ferromagnet-$d$-wave superconductor junctions
}
\author{Jian-Xin Zhu$^{1}$ and C. S. Ting$^{1,2}$
}
\address{ 
$^{1}$Texas Center for Superconductivity and Department of Physics, 
University of Houston, Houston, Texas 77204\\
$^{2}$National Center for Theoretical Sciences, P.O.Box 2-131, Hsinchu,
Taiwan 300, R.O. China}
\maketitle
\begin{abstract}
The proximity effect, quasiparticle 
transport, and local magnetic moment 
in ferromagnet--$d$-wave superconductor junctions with $\{110\}$-oriented
interface are studied by solving self-consistently the Bogoliubov-de 
Gennes equations within an extended Hubbard model. It is found that the 
proximity induced order parameter oscillates in the ferromagnetic region. 
The modulation period is shortened with the increased exchange field while
the oscillation amplitude is depressed by the interfacial scattering. 
With the determined superconducting energy gap, a transfer matrix method 
is proposed to compute the subgap conductance within a scattering approach.
Many novel features including the zero-bias conductance dip and splitting 
are exhibited with appropriate values of the exchange field and  
interfacial scattering strength. 
The conductance spectrum can be influenced seriously by the spin-flip 
interfacial scattering. In addition, a sizable local 
magnetic moment near the $\{110\}$-oriented surface of the $d$-wave 
superconductor is discussed.

\end{abstract} 
\pacs{PACS numbers: 74.20.Mn, 74.80.Fp, 74.50.+r}
]

\narrowtext
\section{Introduction}
The electronic transport in ferromagnetic-superconducting 
hybrid structures is currently a very active area of research due to 
their interesting physical properties and potential device 
applications. A fundamental transport process through the interface 
between the normal conducting and superconducting materials is the Andreev 
reflection (AR):~\cite{Andr64} An electron incident with energy below the 
superconducting energy gap cannot enter the superconductor, it is instead 
reflected at the interface as a hole by transferring a Cooper pair into 
the superconductor. The earlier spin 
polarization experiments involving ferromagnet and superconductor 
were performed on tunnel junctions,~\cite{MT94} where the AR is unimportant 
due to the strong interface barrier. Recently, the effect of 
spin polarization on the AR has been investigated in 
ferromagnet--conventional superconductor 
contacts experimentally,~\cite{UPLB98,SBON98} 
where the AR plays an important role. In an earlier theoretical 
work,~\cite{JB95} this effect was studied in the zero-bias limit.
Several recent spin injection 
experiments~\cite{VLKNG97,DRVJ97,VLNKG98,WYV98} have 
been done with high-$T_c$ superconductors.  
Common to both ferromagnet-conventional and high-$T_c$ superconductor 
junctions, the subgap conductance at a given bias is suppressed due to 
the suppression of AR by the spin splitting of energy band in the 
ferromagnet. In particular, a zero-bias conductance dip was observed 
in the ferromagnet--high-$T_c$ superconductor contacts.~\cite{VLNKG98} 
It has been 
widely accepted that the high-$T_c$ superconductors  
have a $d$-wave pairing symmetry.~\cite{vHarlingen95}
The above interesting observation may  
indicate the importance to take into account the unconventional pairing 
symmetry of the cuprate superconductors. It has been shown that, due to 
the formation of midgap states at the $\{110\}$-oriented 
interface,~\cite{Hu94} the 
conductance spectrum of $d$-wave superconductor junctions differs 
dramatically from that of conventional $s$-wave superconductor 
junctions.~\cite{TK95,XMT96} 
Thus the difference should also exist between ferromagnet--$d$-wave 
and $s$-wave superconductor junctions. In recent theoretical 
works,~\cite{ZFT98,ZO98,KTYB98} the novel features of AR have been 
exhibited in the subgap conductance of 
ferromagnet--$d$-wave superconductor junctions. 
More recently, the effect of spin injection into $s$-
and $d$-wave superconductors has also been studied with an 
emphasis on the interplay between boundary and bulk spin transport 
processes.~\cite{MS99}

In parallel, there also has been much interest 
in the interplay of  superconductivity and ferromagnetism 
in these combined structures.
In the case of ferromagnet--superconductor multilayers,  
the transition temperature of the ($s$-wave) superconductor changes 
nonmonotonically with the thickness or the exchange field strength of the 
ferromagnetic layers.~\cite{Multi1,Multi2,Multi3,Multi4} 
In superconductor--ferromagnet--superconductor
junctions, the exchange field in the ferromagnetic layer leads to 
oscillations of the Josephson critical 
current.~\cite{Multi4,Multi5,DAB97,Note1} 
More recently, the influence of the exchange field on the Josephson 
current in superconductor--ferromagnet--superconductor junctions with 
unconventional pairing symmetry has also been studied.~\cite{TK99} 
In ferromagnet--superconductor--ferromagnet  
multilayers, the appearance of the superconducting energy gap causes 
a reduction of the indirect magnetic 
coupling which exists in the normal state.~\cite{SG95,Melo97}  
For the case of ferromagnet--superconductor junctions, the 
superconducting proximity effect will also change in the presence of  
an exchange field. The previous works~\cite{JB95,ZFT98,ZO98,KTYB98}
with an emphasis on the transport through the ferromagnetic--superconductor 
junctions were based on a simplified continuum model and did not calculate 
the order parameter (i.e., pairing amplitude) self-consistently 
so that the proximity effect cannot be included.  

The purpose of this work is to present a unified and rigorous treatment of 
the proximity effect, transport and magnetic properties in a 
ferromagnet--$d$-wave superconductor junctions.  
Within the framework of an extended Hubbard model, 
we solve self-consistently the Bogoliubov-de Gennes (BdG)
equations to obtain the spatial variation of the order parameter and 
superconducting energy gap. With the obtained energy gap, a transfer matrix 
method is then proposed to calculate the subgap conductance within 
the scattering approach. The self-consistent calculation also allows us 
to study the local magnetic moment in the superconducting region due 
to the presence of the exchange field or Zeeman coupling.  
The main procedure in the present work is parallel to the study of 
transport properties in the normal-metal--anisotropic superconductor 
junctions using the quasiclassical theory, where the pair 
potential first obtained from the quasiclassical formalism is substituted 
into the Andreev equation to calculate the reflection and transmission 
coefficients.~\cite{Bruder90}   

The paper is organized as follows: In Sec.~\ref{SEC:BdG}, the BdG 
equations for the ferromagnet--superconductor junctions are derived within 
the extended Hubbard model. In Sec.~\ref{SEC:OP}, the order parameter and 
pair potential are determined self-consistently. The subgap differential 
conductance and the local magnetic moment are presented in 
Secs.~\ref{SEC:COND} and \ref{SEC:LMM}, respectively. Finally, conclusions 
are given in Sec.~\ref{SEC:SUMMARY}.

\section{The Bogoliubov-de Gennes equations for the 
ferromagnet--superconductor junctions}
\label{SEC:BdG}
We use a single-band extended Hubbard model to describe the 
ferromagnet--superconductor junctions. 
The geometry is shown in Fig.~\ref{FIG:FS} for (a) a ferromagnet--$s$-wave 
superconductor junction  with a $\{100\}$-oriented interface and (b) a 
ferromagnet--$d_{x_{a}^{2}-x_{b}^{2}}$-wave 
superconductor junction with a $\{110\}$-oriented interface. 
The significant difference between $s$-wave and $d$-wave superconductors 
can be exhibited most clearly in these two structures.  
For such a crystalline orientation of 
the $d_{x_{a}^{2}-x_{b}^{2}}$-wave 
superconductor, a $d_{xy}$-wave superconductor junction is formed. 
For the junction involving the $s$-wave superconductor, the qualitative 
features are independent of the interface orientation.  
Hereafter, we will call the ferromagnet--$s$-wave superconductor junction 
with the $\{100\}$-oriented interface the FS junction while the 
ferromagnet--$d_{x_{a}^{2}-x_{b}^{2}}$-wave 
superconductor junction with the $\{110\}$-oriented 
interface the FD$_{xy}$ junction. In the 
junction geometry, both the ferromagnet and the superconductor are treated as 
semi-infinite. Here we choose the interface to be 
at the $0$-th layer. We further assume that the transition 
temperature of the  superconductor is much smaller than the Curie 
temperature of the ferromagnet so that fluctuation effects on the 
magnetism are negligible.  

Under these assumptions, the Hamiltonian defined on two-dimensional square 
lattice is given by 
\begin{eqnarray}
{\cal H}&=&-t\sum_{\langle {\bf ij}\rangle\sigma} 
c_{{\bf i}\sigma}^{\dagger}c_{{\bf j}\sigma}
+\sum_{{\bf i}\sigma}U_{\bf i}n_{{\bf i}\sigma}
+\sum_{{\bf i} \sigma}h_{{\bf i}\sigma}n_{{\bf i}\sigma} 
-\mu\sum_{{\bf i}\sigma}n_{{\bf i}\sigma}
\nonumber \\
&& -\sum_{\bf i}V_{0}({\bf i})n_{{\bf i}\uparrow}
n_{{\bf i}\downarrow}-\frac{1}{2}
\sum_{\langle {\bf ij}\rangle 
\sigma\sigma^{\prime}}V_{1}({\bf ij})n_{{\bf i}\sigma}n_{{\bf 
j}\sigma^{\prime}}\;. \label{EQ:Hubbard}
\end{eqnarray}
Here ${\bf i}$ and ${\bf j}$ are site indices and the angle bracket implies 
that the hopping and interactions are only considered up to 
nearest-neighbor sites, $c_{{\bf i}\sigma}^{\dagger}$ ($c_{{\bf 
i}\sigma}$) are creation (annihilation) operators of an electron with 
spin $\sigma$ on site ${\bf i}$, $n_{{\bf i}\sigma}=c_{{\bf 
i}\sigma}^{\dagger}c_{{\bf 
i}\sigma}$ is the electron number operator on site ${\bf i}$,  
$t$ the hopping integral, and $\mu$ the chemical potential. 
The interfacial scattering potential is modeled by  
$U_{\bf i}=U_{0}\delta_{n0}$,
where $n$ is the layer index along the direction perpendicular to 
the interface plane. The conduction electrons in the 
ferromagnet interact with an 
exchange field, $h_{{\bf i}\sigma}=-h_{0}\sigma_{z}\Theta(-n)$, where 
$\Theta$ is the Heaviside step function and $\sigma_{z}$ $(=\pm 1)$ is the 
eigenvalue of the $z$ component of 
the Pauli matrix. A real space representation of the 
exchange interaction is used since the present system is inhomogeneous. 
The quantities $V_{0}({\bf i})$ 
and $V_{1}({\bf ij})$ are on-site and nearest-neighbor interaction strength, 
respectively. They are taken to be $V_0$ and $V_1$ in the superconductor and 
identically zero in the ferromagnet. Positive values of $V_{0}$ and 
$V_{1}$ mean attractive interactions and negative values mean 
repulsive interactions. When $V_{0}<0$ and $V_{1}>0$, 
the $d$-wave pairing state is  
favorable. Here we also would like to point out that by taking  
the same chemical potential in both the ferromagnet 
and the superconductor, we have ignored the effect of 
the Fermi wavevector mismatch between two materials.  
Very recently, this effect on the conductance spectrum has 
been well studied within the simple continuum model.~\cite{ZO98}     

Within the mean-field approximation, the effective Hamiltonian 
Eq.~(\ref{EQ:Hubbard}) can be written as 
\begin{eqnarray}
{\cal H}_{\mbox{\tiny eff}}&=&-t\sum_{\langle {\bf ij}\rangle\sigma}
c_{{\bf i}\sigma}^{\dagger}c_{{\bf j}\sigma}
+\sum_{{\bf i}\sigma}U_{\bf i}n_{{\bf i}\sigma}
+\sum_{{\bf i} \sigma}h_{{\bf i}\sigma}n_{{\bf i}\sigma} 
-\mu\sum_{{\bf i}\sigma}n_{{\bf i}\sigma}
\nonumber \\
&&+\sum_{\bf i}
[\Delta_{0}^{\dagger}({\bf i})c_{{\bf i}\downarrow}c_{{\bf i}\uparrow}
+\Delta_{0}({\bf i})c_{{\bf i}\uparrow}^{\dagger}c_{{\bf i}
\downarrow}^{\dagger}]
\nonumber \\
&&+\sum_{\langle {\bf i},{\bf j}
={\bf i}+\mbox{\boldmath{$\delta$}}\rangle } 
[\Delta_{\mbox{\boldmath{$\delta$}}}^{\dagger}({\bf i})
c_{{\bf j}\downarrow}c_{{\bf i}\uparrow}
+\Delta_{\mbox{\boldmath{$\delta$}}}({\bf i})
c_{{\bf i}\uparrow}^{\dagger}c_{{\bf j}\downarrow}^{\dagger}] 
\;,
\label{EQ:Hubbard1}
\end{eqnarray}
where 
\begin{equation}
\Delta_{0}({\bf i})=V_{0}({\bf i})\langle c_{{\bf i}\uparrow}
c_{{\bf i}\downarrow}\rangle\;, 
\end{equation} 
\begin{equation}
\Delta_{\mbox{\boldmath{$\delta$}} }
({\bf i})=V_{1}({\bf i},{\bf i}+\mbox{\boldmath{$\delta$}})
\langle c_{{\bf 
i}\uparrow} c_{{\bf i}+\mbox{\boldmath{$\delta$}}
\downarrow}\rangle\;,
\end{equation}
are the on-site and nearest-neighbor pair potentials, respectively.
The effective Hamiltonian Eq.~(\ref{EQ:Hubbard1}) can be diagonalized as 
$H_{\mbox{\tiny eff}}=E_{g}+\sum_{\nu} 
E_{\nu}\gamma_{\nu}^{\dagger}\gamma_{\nu}$ by 
performing the Bogoliubov transformation, 
\begin{equation}
c_{{\bf i}\sigma}=\sum_{\nu}[u_{{\bf i}\sigma}^{\nu}\gamma_{\nu}
-\sigma v_{{\bf i}\sigma}^{\nu*}\gamma_{\nu}^{\dagger}]\;.
\end{equation}
Here $\gamma_{\nu}^{\dagger}$ ($\gamma_{\nu}$) are creation
(annihilation) operators of fermionic quasiparticles. 
$E_{\nu}$ are the quasiparticle eigenvalues. The 
quasiparticle wavefunction amplitudes
$(u_{{\bf i}\sigma},v_{{\bf i}\sigma})$  
satisfy the lattice BdG equations~\cite{deG66}
\begin{mathletters}
\label{EQ:BdG}
\begin{eqnarray}
\label{EQ:BdGa}
&\sum_{\bf j} \left( \begin{array}{cc} 
H_{\bf ij}+h_{\bf i}\delta_{\bf ij} & \Delta_{\bf ij} \\
\Delta_{\bf ij}^{\dagger} & -(H_{\bf ij}-h_{\bf i}\delta_{\bf ij}) 
\end{array} \right) \left( \begin{array}{c}
u_{{\bf j}\uparrow}^{\nu} \\ v_{{\bf j}\downarrow}^{\nu} 
\end{array} \right) & \nonumber \\
&=E_{\nu} \left( \begin{array}{c}
u_{{\bf i}\uparrow}^{\nu} \\ v_{{\bf i}\downarrow}^{\nu} 
\end{array} \right)\;, &
\end{eqnarray}
\begin{eqnarray}
\label{EQ:BdGb}
&\sum_{\bf j} \left( \begin{array}{cc} 
H_{\bf ij}-h_{\bf i}\delta_{\bf ij} & \Delta_{\bf ij} \\
\Delta_{\bf ij}^{\dagger} & -(H_{\bf ij}+h_{\bf i}\delta_{\bf ij}) 
\end{array} \right) \left( \begin{array}{c}
u_{{\bf j}\downarrow}^{\nu} \\ v_{{\bf j}\uparrow}^{\nu} 
\end{array} \right) & \nonumber \\
&=E_{\nu} \left( \begin{array}{c}
u_{{\bf i}\downarrow}^{\nu} \\ v_{{\bf i}\uparrow}^{\nu} 
\end{array} \right)\;, &
\end{eqnarray}
\end{mathletters}
where
\begin{equation}
H_{\bf ij}=-t\delta_{{\bf i}+\mbox{\boldmath{$\delta$}},{\bf j}}
+(U_{\bf i}-\mu)\delta_{\bf ij}\;,
\end{equation} 
\begin{equation}
\Delta_{\bf ij}=\Delta_{0}({\bf i})\delta_{\bf ij}
+\Delta_{\mbox{\boldmath{$\delta$}}}({\bf i})
\delta_{{\bf i}+\mbox{\boldmath{$\delta$}},{\bf j}}\;,
\end{equation}
with $\mbox{\boldmath{$\delta$}}=\pm \hat{\bf x}_{a},
\pm \hat{\bf x}_{b}$ the unit vectors along the crystalline $x_{a}$ 
and $x_{b}$ axis.
The energy gaps for on-site and nearest-neighbor pairing are 
determined self-consistently     
\begin{eqnarray}
\Delta_{0}({\bf i})&=&V_{0}({\bf i})F_{0}({\bf i}) \nonumber \\
&=&\frac{V_{0}({\bf i})}{2}\sum_{\nu} [u_{{\bf i}\uparrow}^{\nu}v_{{\bf 
i}\downarrow}^{\nu*} +u_{{\bf i}\downarrow}^{\nu}v_{{\bf i}\uparrow}^{\nu*}] 
\tanh (E_{\nu}/2T)\;,
\label{EQ:Gap0}
\end{eqnarray}
\begin{eqnarray}
\Delta_{\mbox{\boldmath{$\delta$}}} ({\bf i})
&=&V_{1}({\bf i},{\bf i}+\mbox{\boldmath{$\delta$}})
F_{\mbox{\boldmath{$\delta$}}} ({\bf i}) \nonumber \\
&=&\frac{V_{1}({\bf i},{\bf i}+\mbox{\boldmath{$\delta$}})}{2}
\sum_{\nu}[u_{{\bf i}\uparrow}^{\nu}v_{
{\bf i}+\mbox{\boldmath{$\delta$}}, \downarrow}^{\nu*}
+u_{{\bf i}+\mbox{\boldmath{$\delta$}},\downarrow}^{\nu}v_{{\bf 
i}\uparrow}^{\nu*}] \nonumber \\
&&\times \tanh (E_{\nu}/2T)\;, 
\label{EQ:Gap1}
\end{eqnarray}
where the Boltzmann constant $k_{B}=1$ has been taken, and $F_{0}({\bf i})$ 
and $F_{\mbox{\boldmath{$\delta$}}}({\bf i})$ are the on-site and 
nearest-neighbor bond order parameter.
Note that the $4\times 4$ BdG equations are decoupled into two sets 
of $2 \times 2$ equations since the spin-flip effect is not considered in 
Eq.~(\ref{EQ:Hubbard}).
Note that the eigenstates of the BdG equations exist in pairs: 
{\em If $(u_{\uparrow},v_{\downarrow})$ is the solution of 
Eq.~(\ref{EQ:BdGa}) with the eigenvalue $E$, then 
$(-v_{\downarrow}^{*},u_{\uparrow}^{*})$ is 
the solution of Eq.~(\ref{EQ:BdGb}) with the eigenvalue $-E$}. 
  
For a clean ferromagnet-superconductor junction with a flat interface,
which we are considering, 
there exists the translation symmetry along the $y$ direction so that  
the Bloch theorem can be applied to this direction.
Then for the FD$_{xy}$ junction, the eigenfunction can be written 
in the form 
\begin{equation}
\left( \begin{array}{c}
u_{{\bf i}=n,m}\\ v_{{\bf i}=n,m} 
\end{array}\right) =\frac{1}{\sqrt{N_y}}
\left( \begin{array}{c}
u_{n}(k_{y})\\ v_{n}(k_{y}) 
\end{array} \right) e^{imk_{y}a/\sqrt{2}}\;,
\end{equation}
where $a$ is the lattice constant, 
$k_{y}\in [-\pi/\sqrt{2}a,\pi/\sqrt{2}a]$, $m$ and $n$ are  
ionic-layer indices in the $x$ and $y$ directions, and $N_y$ is the number 
of unit cells along the $y$ direction.  
The problem becomes solving the BdG equations for 
$(u_{n}(k_{y}),v_{n}(k_{y}))$ corresponding to the eigenvalue 
$E(k_y)$:
\begin{mathletters}
\begin{eqnarray}
\label{EQ:BdGa-1D}
&\sum_{\bf n^{\prime}} \left( \begin{array}{cc} 
H_{nn^{\prime}}+h_{n}\delta_{nn^{\prime}} & \Delta_{nn^{\prime}} \\
\Delta_{nn^{\prime}}^{\dagger} & -(H_{nn^{\prime}}
-h_{n}\delta_{nn^{\prime}}) 
\end{array} \right) \left( \begin{array}{c}
u_{n^{\prime}\uparrow}^{\nu} \\ v_{n^{\prime}\downarrow}^{\nu} 
\end{array} \right) & \nonumber \\
& =E_{\nu}(k_{y}) \left( \begin{array}{c}
u_{n\uparrow}^{\nu} \\ v_{n\downarrow}^{\nu} 
\end{array} \right)\;, &
\end{eqnarray}
\begin{eqnarray}
\label{EQ:BdGb-1D}
& \sum_{n^{\prime}} \left( \begin{array}{cc} 
H_{nn^{\prime}}-h_{n}\delta_{nn^{\prime}} & \Delta_{nn^{\prime}} \\
\Delta_{nn^{\prime}}^{\dagger} & 
-(H_{nn^{\prime}}+h_{n}\delta_{nn^{\prime}}) 
\end{array} \right) \left( \begin{array}{c}
u_{n^{\prime}\downarrow}^{\nu} \\ v_{n^{\prime}\uparrow}^{\nu} 
\end{array} \right) & \nonumber \\
& =E_{\nu}(k_y) \left( \begin{array}{c}
u_{n\downarrow}^{\nu} \\ v_{n\uparrow}^{\nu} 
\end{array} \right)\;, &
\end{eqnarray}
\end{mathletters} 
where 
\begin{equation}
H_{nn^{\prime}}=-2t\cos (k_{y}a/\sqrt{2}) \delta_{ 
n\pm 1,n^{\prime}}
+(U_{n}-\mu)\delta_{nn^{\prime}}\;,
\end{equation} 
\begin{eqnarray}
\Delta_{nn^{\prime}}&=&\Delta_{0}(n)\delta_{nn^{\prime}}
+[\Delta_{a}(n,n\pm 1)e^{\mp ik_y a/\sqrt{2}}
\nonumber \\
&& +\Delta_{b}(n,n\pm 1)e^{\pm ik_y a/\sqrt{2}}]
\delta_{n\pm 1,n^{\prime}}\;,
\end{eqnarray}
with the gap functions given by 
\begin{equation}
\Delta_{0}(n)
=\frac{V_{0}(n)}{2N_y}\sum_{\nu,k_y} [u_{n\uparrow}^{\nu}
v_{n\downarrow}^{\nu*}+u_{n\downarrow}^{\nu}
v_{n\uparrow}^{\nu*}] \tanh [E_{\nu}(k_y)/2T]\;,
\label{EQ:Gap0-1D}
\end{equation}
\begin{eqnarray}
\Delta_{a}(n,n\pm 1)
&=&\frac{V_{1}(n,n\pm 1)}{2N_y}\sum_{\nu,k_y}[u_{n\uparrow}^{\nu}v_{
n\pm 1, \downarrow}^{\nu*}e^{\pm i k_{y}a/\sqrt{2}}
\nonumber \\ 
&& +u_{n\pm 1,\downarrow}^{\nu}v_{n \uparrow}^{\nu*}
e^{\mp i k_{y} a/\sqrt{2}}
] \nonumber \\
&& \times
\tanh [E_{\nu}(k_y)/2T]\;, 
\label{EQ:Gap1a-1D}
\end{eqnarray}
and 
\begin{eqnarray}
\Delta_{b}(n,n\pm 1)
&=&\frac{V_{1}(n,n\pm 1)}{2N_y}\sum_{\nu,k_y}[u_{n\uparrow}^{\nu}v_{
n\pm 1, \downarrow}^{\nu*}e^{\mp i k_{y}a/\sqrt{2}}
\nonumber \\
&&+u_{n\pm 1,\downarrow}^{\nu}v_{n \uparrow}^{\nu*}
e^{\pm i k_{y} a/\sqrt{2}}
] \nonumber \\
&& \times
\tanh [E_{\nu}(k_y)/2T]\;. 
\label{EQ:Gap1b-1D}
\end{eqnarray}
The problem with other orientations of the flat interface can be treated 
similarly.  
 
\section{Self Determination of the Order Parameter and the Pair Potentials} 
\label{SEC:OP}

We solve the BdG equations self-consistently by starting with an 
initial gap function. After exactly diagonalizing 
Eq.~(\ref{EQ:BdG}), the obtained Bogoliubov amplitudes are  
substituted into Eqs.~(\ref{EQ:Gap0}) and (\ref{EQ:Gap1}) to 
compute a new gap function. We then use it as input to repeat the above 
process until the relative error in the gap function between successive 
iterations is less than the desired accuracy. 
Throughout this work, we concentrate on the zero temperature case
unless specified explicitly, and take 
the parameters: $\mu=0$, and $V_{1}=2t$ and $V_{0}=-2t$ for $d$-wave 
superconductors while $V_{1}=0$ and $V_{0}=2t$ for $s$-wave superconductors. 
This set of parameter values give 
the zero-temperature energy gap  
$\Delta_{d0}=0.241t$ 
and $\Delta_{s0}=0.376t$ 
for the bulk $d$-wave and $s$-wave superconductors, respectively. 
The corresponding coherence $\xi_{d}\approx 1.3a$ and $\xi_{s}
\approx 3.4a$. 
Note that as in other works,~\cite{SKB94,FKB96} the model 
parameters chosen here are not intended for realistic materials.
    
For the $d$-wave superconductor, the amplitudes of $d$- and extended 
$s$-wave order parameters 
can be defined in terms of the bond order parameters:~\cite{SKB94} 
\begin{mathletters}
\begin{eqnarray}
F_{d}({\bf i})&=&\frac{1}{4}[F_{\hat{x}_a}({\bf i})
+F_{-\hat{x}_a}({\bf i})
-F_{\hat{x}_b}({\bf i})
-F_{-\hat{x}_b}({\bf i})]\;,  \\
F_{s}({\bf i})&=&\frac{1}{4}[F_{\hat{x}_a}({\bf i})+
F_{-\hat{x}_a}({\bf i})
+F_{\hat{x}_b}({\bf i})
+F_{-\hat{x}_b}({\bf i})]\;. 
\end{eqnarray}
\end{mathletters}
Accordingly, the $d$-wave and extended $s$-wave pair potentials are given 
by: 
\begin{mathletters}
\begin{eqnarray} 
\Delta_{d}({\bf i})&=&\frac{1}{4}[\Delta_{\hat{x}_a}({\bf i})
+\Delta_{-\hat{x}_a}({\bf i})
-\Delta_{\hat{x}_b}({\bf i})
-\Delta_{-\hat{x}_b}({\bf i})]\;,  \\
\Delta_{s}({\bf i})&=&\frac{1}{4}[\Delta_{\hat{x}_a}({\bf i})+
\Delta_{-\hat{x}_a}({\bf i})
+\Delta_{\hat{x}_b}({\bf i})
+\Delta_{-\hat{x}_b}({\bf i})]\;.
\end{eqnarray}
\end{mathletters}
In the superconducting region, the energy gap is proportional to the 
order parameter because of the constant pairing interaction.
In the bulk state of $d$-wave superconductor, the extended $s$-wave 
component is zero. For the junction systems under consideration, 
the induced extended $s$-wave component near the interface 
is numerically found to be 
vanishingly small for the value of on-site repulsive interaction we have 
taken. For the conventional $s$-wave superconductor, 
the order parameter and the pair potential are directly on-site defined.

In Figures~\ref{FIG:OP-SH} and \ref{FIG:OP-DH}, we plot the spatial 
variation of the order parameter for various values of exchange field in 
the FS junction and the FD$_{xy}$ junction. 
In this case, there is no interfacial scattering potential.  
As can be seen, the exchange field does not influence the order parameter 
in the superconducting region. 
In the normal metal case ($h_{0}=0$), the proximity 
induced order parameter monotonically decays into the 
normal metal region. Interestingly, common to both the FS and FD$_{xy}$ 
junctions, the order parameter in the ferromagnetic region 
no longer changes monotonically, it instead oscillates around the zero 
value of order parameter. In addition, as the exchange field is 
increased, the oscillation period becomes shorter. This behavior could be 
understood in the 
following way. Take the FS junction as an example. 
Since the component of the wavevector 
parallel to the interface is conserved, 
we can just consider the normal component.
In the superconducting 
region, the wavevectors (momenta) 
of the spin-up and spin-down electrons forming the Cooper pairs 
have the same amplitude (but opposite directions) $q_x$.  
Upon entering into the ferromagnetic region, the pair amplitude decays. 
Simultaneously, the spin-up electron lowers its energy by $h_{0}$, while 
the spin-down electron gain the energy $h_{0}$. In order for each 
electron to conserve its total energy, the spin-up 
electron should adjust its momentum to $q_{\uparrow}$, while the 
spin-down electron to $q_{\downarrow}$. Therefore, from the 
expressions of the order parameter given by~(\ref{EQ:Gap0}), 
we can approximately write the order parameter as 
$\cos[(q_{\uparrow}-q_{\downarrow})na]\Phi(n)$, where $\Phi(n)$ 
is a slow-varying envelope function. Therefore, the exchange field in the 
ferromagnet causes the spatial modulation of the order parameter, which now 
roughly varies at the scale of $(q_{\uparrow}-q_{\downarrow})^{-1}$. 
In the continuum model, 
the difference $q_{\uparrow}-q_{\downarrow}\approx 2h_{0}/\hbar v_{Fx}$, 
where $v_{Fx}$ is the normal component of the Fermi velocity.     
This also explains the decrease of the modulation period with the exchange 
field $h_{0}$. 
In Figs.~\ref{FIG:OP-SI} and \ref{FIG:OP-DI}, the spatial variation of order 
parameter are plotted for various values of the 
interfacial scattering potential in the FS and FD$_{xy}$ junctions 
with the exchange field fixed at $h_{0}=0.125D$ ($D=8t$ is the band 
width). Our numerical results show that as the interfacial scattering potential 
becomes stronger, the oscillation amplitude of the order parameter in the 
ferromagnet is decreased. This is because the amplitude of the slow-varying 
envelope function mentioned above has been suppressed by the interfacial 
scattering at the interface. However, in
the superconducting region, the order parameters of the FS and FD$_{xy}$ 
junctions show different behavior in the presence of the interfacial 
scattering. As shown in Figs.~\ref{FIG:OP-SI}, for the FS junction, the 
depression of the order parameter near the interface 
is decreased by the interfacial scattering. 
As the interface is strongly reflecting (large $U_{0}$), the 
superconductor and the ferromagnet are almost decoupled, and since  
the opaque interface itself is not pair breaking for $s$-wave 
superconductivity, the $s$-wave order parameter is not depressed.       
In contrast to the $s$-wave case, in the FD$_{xy}$ junction,  
the reflected quasiparticles from the $\{110\}$-oriented  interface are 
subject to a sign change of the order parameter, 
which makes the opaque interface 
itself pair-breaking. Thus the $d$-wave order parameter is strongly 
depressed (see Fig.~\ref{FIG:OP-DI}). Finally, since we have assumed 
that there is no pairing interaction in the ferromagnet, the pair 
potential or energy gap in this region is zero. It is the pair potential 
that acts as an off-diagonal scattering potential in the BdG 
equations.  

\section{The subgap differential conductance}
\label{SEC:COND}
\subsection{The case of nonmagnetic interfacial scattering}
Once the BdG equations~(\ref{EQ:BdG}) are solved self-consistently, we 
can use the obtained pair potential  to calculate the differential 
conductance. The transport properties 
through the normal-metal--superconductor junctions can be studied  
within the  Blonder-Tinkham-Klapwijk (BTK) scattering formalism,~\cite{BTK82} 
which  expresses the differential conductance in terms of the normal and 
Andreev reflection 
coefficients. In contrast to the tunneling Hamiltonian model, which 
requires an opaque barrier at the interface, the BTK theory can consider 
the case of an arbitrary barrier strength. Also noticeably, the BTK 
formalism can be regarded as the earliest version of the 
Landauer-B\"{u}ttiker~\cite{LB70-86} formula 
applied to the coherent transport
through a normal-metal--superconductor 
structure.~\cite{Beenakker91} 
Recently, the BTK theory has been extended 
to the spin-dependent transport 
through ferromagnet-superconductor junctions.~\cite{JB95} 
Within the tight-binding model, the averaged differential conductance
can be obtained as: 
\begin{equation}
G=\frac{e^{2}}{h N_{y}}\sum_{k_y,\sigma}
[1+R_{h,\underline{\sigma}\sigma}-R_{e,\sigma\sigma}]
\;,
\label{EQ:COND-AV}
\end{equation}
which shows clearly that an incoming electron of spin
$\sigma$ ($=\uparrow,\downarrow$) is normally reflected as an electron of 
the same spin $\sigma$ with probability
$R_{e,\sigma\sigma}=\vert r_{\sigma\sigma}\vert^{2}$,
and Andreev reflected as a hole of the opposite spin $\underline{\sigma}$
with probability $R_{h,\underline{\sigma}\sigma}=[\sin
(q_{\underline{\sigma}}a)/\sin (q_{\sigma}a)]\vert
r_{\underline{\sigma}\sigma} \vert^{2}$.
Here the summation is over all the transverse modes and over the spin 
indices. Without confusion, $h$ in Eq.~(\ref{EQ:COND-AV}) is the Planck 
constant. In contrast to the continuum model, the factor 
$\sin (q_{\underline{\sigma}}a)/\sin (q_{\sigma}a)$ comes from the band 
structure effect. 
Our previous work within the 
continuum model~\cite{ZFT98} concentrated on the direction-dependent 
subgap conductance through the ferromagnet--superconductor junctions, 
which can be experimentally explored with the scanning tunneling 
spectroscopy.~\cite{WYGS98} 
For a point contact junction,~\cite{VLNKG98} 
a summation over the transverse modes is needed. 
   
The remaining thing is to obtain the Andreev and normal reflection 
coefficients, which can be calculated using the transfer matrix 
method. As an illustration, we  
give a detailed procedure for the calculation of these coefficients for 
the FS junction, which has a $\{100\}$-oriented interface.
From the BdG equations~(\ref{EQ:BdG}), we can write 
the relation of wavefunctions among consecutive layers: 
\begin{equation}
\left(
\begin{array}{c}
u_{n+1} \\
v_{n+1} \\
u_{n} \\
v_{n}
\end{array} 
\right) = \hat{\bf M}_{n} 
\left(
\begin{array}{c}
u_{n} \\ v_{n} \\ u_{n-1} \\ v_{n-1}
\end{array} 
\right)
\end{equation}
where the transfer matrix for $n$-th layer is given by 
\begin{equation}
\hat{\bf M}_{n}=\left( 
\begin{array}{cccc}
\frac{\tilde{\epsilon}_{n}-h_{n}-E}{t} & \frac{\Delta_{s}(n)}{t} & -1 & 0 \\
-\frac{\Delta_{s}^{*}(n)}{t} & \frac{\tilde{\epsilon}_{n}+h_{n}+E}{t} 
& 0 & -1 \\
1& 0 & 0 & 0\\
0 & 1 & 0 &0
\end{array}
\right)\;,
\end{equation}
with $\tilde{\epsilon}_{n}=U_{0}\delta_{n0}-2t\cos k_{y}a-\mu$ 
and $h_{n}=h_0 \Theta(-n)$. 
The incident and transmitted wave amplitudes are then connected with 
the total transfer matrix
\begin{equation}
\left(
\begin{array}{c}
u_{N_{R}+1} \\
v_{N_{R}+1} \\
u_{N_{R}} \\
v_{N_{R}}
\end{array} 
\right) = \hat{\bf M}_{T} 
\left(
\begin{array}{c}
u_{N_{L}} \\ v_{N_{L}} \\ u_{N_{L}-1} \\ v_{N_{L}-1}
\end{array} 
\right)\;,
\label{EQ:COE}
\end{equation}
where
\begin{equation}
\hat{\bf M}_{T}=\prod_{n} \hat{\bf M}_{n}\;.
\end{equation}
Here $N_{L,R}$ are the indices for two outmost layers of the scattering 
region where the energy gap has approached its bulk value. If the electron 
wave is incident with spin up 
and transverse momentum $k_{y}$, the incident, reflected and transmitted 
wavefunctions are written as
\begin{mathletters}
\begin{equation}
\left(
\begin{array}{c}
u_{N_{L}} \\
v_{N_{L}}
\end{array} 
\right)=
\left(
\begin{array}{c}
1 \\
0 
\end{array} 
\right)
+r_{\downarrow\uparrow}\left(
\begin{array}{c}
0 \\
1 
\end{array} 
\right)
+r_{\uparrow\uparrow}\left(
\begin{array}{c}
1\\
0
\end{array} 
\right)\;, 
\end{equation}
\begin{eqnarray}
\left(
\begin{array}{c}
u_{N_{L}-1} \\
v_{N_{L}-1}
\end{array} 
\right) &=&
\left(
\begin{array}{c}
1 \\
0 
\end{array} 
\right)e^{-iq_{\uparrow}a}
+r_{\downarrow\uparrow}\left(
\begin{array}{c}
0 \\
1 
\end{array} 
\right)e^{-iq_{\downarrow}a}
\nonumber \\
&& +r_{\uparrow\uparrow}\left(
\begin{array}{c}
1\\
0
\end{array} 
\right)e^{iq_{\uparrow}a}\;, 
\end{eqnarray}   
\end{mathletters}
and 
\begin{mathletters}
\begin{equation}
\left(
\begin{array}{c}
u_{N_{R}} \\
v_{N_{R}}
\end{array} 
\right)=
t_{\uparrow\uparrow}
\left(
\begin{array}{c}
u_{+}e^{i\phi_{+}} \\
v_{+}  
\end{array} 
\right)+
t_{\downarrow\uparrow}
\left(
\begin{array}{c}
v_{-}e^{i\phi_{-}} \\
u_{-}  
\end{array} 
\right)\;, 
\end{equation}
\begin{eqnarray}
\left(
\begin{array}{c}
u_{N_{R}+1} \\
v_{N_{R}+1}
\end{array} 
\right) &=&
t_{\uparrow\uparrow}
\left(
\begin{array}{c}
u_{+}e^{i\phi_{+}} \\
v_{+}  
\end{array} 
\right)e^{ik_{e}a}
+t_{\downarrow\uparrow}
\left(
\begin{array}{c}
v_{-}e^{i\phi_{-}} \\
u_{-}  
\end{array} 
\right)\nonumber \\
&&\times e^{-ik_{h}a}\;. 
\end{eqnarray}    
\end{mathletters}
Here $t_{\uparrow\uparrow}$ and $t_{\downarrow\uparrow}$ are the 
transmission amplitudes. $\phi_{\pm}$ are the internal phase of the 
energy gap and are identically zero for the $s$-wave superconductor.
The wave vectors in the ferromagnet and the superconductor are, respectively,
\begin{equation}
q_{\uparrow,\downarrow}a=\cos^{-1}\left[ -\frac{2t\cos k_{y}a+\mu \pm 
(E+h_{0})}{2t} \right],
\end{equation}
and
\begin{equation}
k_{e,h}a=\cos^{-1}\left[ -\frac{2t\cos k_{y}a+\mu \pm
\sqrt{E^{2}-\vert \Delta({\bf k}_{\pm})\vert^{2}} }{2t}
\right]\;.
\end{equation}
The BCS coherence factors are given by 
\begin{mathletters}
\begin{eqnarray}
u_{\pm}^{2}&=&\frac{1}{2}\left[ 1+
\frac{ \sqrt{E^{2}-\vert \Delta({\bf k}_{\pm})\vert^{2} }}{E}\right]\;,\\
v_{\pm}^{2}&=&\frac{1}{2}\left[ 1-
 \frac{ \sqrt{E^{2}-\vert \Delta({\bf k}_{\pm})\vert^{2}}}{E}\right]\;.
\end{eqnarray} 
\end{mathletters}
For clarity, we have written the energy gap explicitly depending on the 
wavevector $\Delta({\bf k}_{\pm})\equiv \Delta(\pm q_{0},k_y)$ with 
the $x$-component of the Fermi wavevector $q_{0}a=\cos^{-1}[-(2t\cos k_y a 
+\mu)/2t]$. For a conventional $s$-wave superconductor, 
$\Delta({\bf k}_{\pm})\equiv \Delta_{s0}$.
Then the reflection amplitudes can be obtained by solving the 
linear equation~(\ref{EQ:COE}).
The reflection coefficients for the FD$_{xy}$ junction can be 
calculated similarly, where $q_{0}a/\sqrt{2}=\cos^{-1}[-\mu/4t\cos (k_y 
a/\sqrt{2})]$ and $\Delta({\bf k}_{+})=-\Delta({\bf k}_{-})
=4\Delta_{d0} \sin(q_0a/\sqrt{2})
\sin(k_y a/\sqrt{2})$ so that the internal phase $\phi_{+}=0$ (or $\pi$) 
while $\phi_{-}=\pi$ (or $0$) depending on 
$\Delta({\bf k}_{+})$ being positive or negative. 
Note that the reflection coefficients,
which we are calculating here, can also 
be used to study the conductance spectrum 
for the spin current.~\cite{KTYB98} 

Before presenting the results for the conductance, we give a physical 
analysis of the effect of exchange field on the AR. In     
Fig.~\ref{FIG:AR},  we 
schematically draw the spin-split energy band in the ferromagnet and the 
AR process in the continuum model. As shown in the figure, 
the incident electrons and the Andreev reflected holes occupy different 
spin bands. Thus the AR is sensitive to the relative spin-dependent 
density of states at the Fermi energy $E_{F}$. In the normal metal 
($h_0=0$), the energy band is spin degenerate, the AR is thus not 
suppressed. However, if the exchange field is sufficiently strong that 
there are no electrons occupying the spin-down band in the ferromagnet, 
the incident spin-up electrons have no spin-down electrons to drag in 
order to form Cooper pairs. As a consequence, the AR is completely 
depressed. For the numerical calculation, we take $N_y=625$. In 
Figs.~\ref{FIG:FS0} and \ref{FIG:FD0}, 
the subgap conductance spectrum $G$ 
versus the scaled energy $E$ is plotted for the FS junction 
and the FD$_{xy}$ junction with various values of $h_{0}$ 
but without the interfacial scattering. 
As can be seen, for both the FS and FD$_{xy}$ junctions, the averaged 
conductance at a given energy $E$ is suppressed due to the blocking of AR.
However, because the effective energy gap for the $d$-wave pairing 
symmetry is momentum-dependent while that of $s$-wave pairing symmetry is 
a constant in the momentum space, the 
different conductance behaviors between  the FS and 
FD$_{xy}$ junctions are exhibited. 
In the FS junction, before the subgap conductance is completely 
suppressed, a flat zero-bias maximum always shows up in the conductance 
spectrum.  When the exchange field is sufficiently strong, the 
conductance is zero within the energy gap, and sharply increases to a 
finite value as the bias crosses the gap edge. This is because, outside the 
energy gap, the normal conduction process becomes important.  
In the FD$_{xy}$ junction, as the 
exchange field becomes strong, a zero-bias conductance maximum 
gives way to a zero-bias conductance dip. 
The strikingly similarity between the lowest curve in 
Fig.~\ref{FIG:FD0} and the experimental measurement 
performed on the  
La$_{2/3}$Ba$_{1/3}$MnO$_{3}$/DyBa$_{2}$Cu$_{3}$O$_{7}$ 
junctions~\cite{VLNKG98} 
demonstrates that the high degree of spin 
polarization in the doped lanthanum manganite 
compounds and the $d$-wave pairing symmetry of the high-$T_{c}$ 
superconductors are essential to explain the observed 
conductance behavior. In addition, as is shown,   
the conductance spectrum in the ferromagnet--superconductor 
junction is symmetric to the zero bias, which has been observed in many 
experiments on spin polarized transport.~\cite{UPLB98,SBON98,VLNKG98}
A general proof of this property is given in the Appendix~\ref{SEC:APPD}.

In Figs.~\ref{FIG:FSI} and \ref{FIG:FDI}, we plot the conductance spectrum 
for a variety values of exchange field with the 
barrier strength fixed at $U_{0}=0.2D$ in the FS junction and 
$U_{0}=0.625D$ in 
the FD$_{xy}$ junction. In this case, the overall conductance spectrum is 
also reduced by the increase of $h_{0}$. In the FS junction, a gap-like 
structure is exhibited in the conductance, and the peak at the gap edge 
is remarkably depressed by the exchange field, which is consistent with 
the recent experimental observations on ferromagnet--$s$-wave 
superconductor junctions where the degree of spin polarization is small 
and a small barrier scattering potential 
may still exist.~\cite{UPLB98} 
In the FD$_{xy}$ junction, due to the existence of midgap 
states at the interface, a sharp zero-bias conductance peak (ZBCP) shows up. 
The amplitude of this conductance peak is strongly suppressed by the 
exchange field. Meanwhile, as the exchange field becomes much 
stronger, the highly suppressed ZBCP is split. 
Physically, the suppression of the $d$-wave order parameter near 
the interface allows the ferromagnetic effect to penetrate into the 
superconductor side through the tunneling of electrons, which leads to a 
small imbalance of the local occupation of electron with different spin 
direction so that a small magnetization at the $d$-wave superconductor 
side is induced.  The small magnetization in turn causes the shift of 
the energy of the midgap states and the conductance peak is split.    
This splitting depends on the transparency of the interface. For a very 
strong barrier, the splitting is almost unobservable.  
The splitting of the ZBCP by the exchange interaction can 
also be realized by the application of a magnetic field. If an in-plane 
magnetic field $B$ is applied parallel to the interface,  
the orbital coupling between electrons and the magnetic field 
can be neglected and only the Zeeman coupling $\mp\mu_{B}B$ 
($\mu_{B}$ is the Bohr magneton) is present.  
Unlike the ferromagnetic effect on the electronic structures in the 
superconducting region near the interface, which is essentially of 
dynamic origin, the Zeeman coupling is purely a local interaction. In 
this situation, the electron energy globely shifts to $E\pm \mu_{B} B$ so 
that the energy of the midgap states shifts $\mu_{B}B$. 
Consequently, as shown in Fig.~\ref{FIG:FDB}, the ZBCP in the 
normal-metal--$d_{x_{a}^{2}-x_{b}^{2}}$-wave superconductor junction with 
$\{110\}$-oriented  interface (From now on we call it the ND$_{xy}$ 
junction) can be readily split. The range of splitting is just $2\mu_{B} B$.

\subsection{The effects of spin-flip interfacial scattering}
In the preceding treatment, the spin-flip interfacial scattering effects
are ignored. In case of the junctions with the ferromagnet involved, 
this type of scattering may be important. To study the effect,
we introduce a new term into the Hamiltonian
\begin{equation}
{\cal H}_{sp}= \sum_{{\bf i},\sigma \neq \sigma^{\prime}}
U_{{\bf i},\sigma\sigma^{\prime}}^{sp}
c_{{\bf i}\sigma}^{\dagger}c_{{\bf
i}\sigma^{\prime}}\;,
\end{equation}
where $U_{{\bf
i},\sigma\sigma^{\prime}}^{sp}=U_{1}\delta_{0n}
\delta_{\underline{\sigma}\sigma^{\prime}}$ is assumed to be nonzero
at the interface (layer index $n=0$).
In the spin-space, the spin-flip scattering term is represented by
\begin{equation}
\hat{U}^{sp}=\left( \begin{array}{cc} 
0 &U_{1} \\
U_{1} & 0 \\
\end{array} \right)\;.
\end{equation}
The spin-nonflip term has been represented by $U_{\bf i}$ in the total
Hamiltonian, which is the diagonal elements in the spin space.
In the normal state junction with the spin-flip scattering, a beam of 
incident
electrons with spin $\sigma$ will be reflected as electrons with the
same spin and opposite spin.
In the junction made up of the superconductors, one can expect that, when a
beam of electrons with spin $\sigma$ incident from the normal metal or
ferromagnet, the spin-flip scattering will also lead to
the normally reflected electrons with the
opposite spin $\underline{\sigma}$ and Andreev reflected holes
with the same spin $\sigma$, in addition to
the reflected electrons with the same spin $\sigma$
and holes with the opposite spin $\underline{\sigma}$.
By working with the coupled $4\times 4$ BdG matrix equations,
we can generalize the previous transfer matrix technique to obtain
the normal reflection amplitudes,
$r_{e,\sigma\sigma}$, $r_{e,\underline{\sigma}\sigma}$, and the Andreev
reflection amplitudes, $r_{h,\underline{\sigma}\sigma}$, 
$r_{h,\sigma\sigma}$.
Correspondingly, the conductance is generalized to be
\begin{equation}
G=\frac{e^{2}}{h N_{y}}\sum_{k_y,\sigma}   
[1+R_{h,\underline{\sigma}\sigma}
+R_{h,\sigma\sigma}
-R_{e,\sigma\sigma}
-R_{e,\underline{\sigma}\sigma}]
\;,
\label{EQ:COND-AV1}
\end{equation}
where $R_{h,\underline{\sigma}\sigma}
=[\sin (q_{h,\underline{\sigma}}a)/\sin (q_{e,\sigma}a)]\vert
r_{h,\underline{\sigma}\sigma} \vert^{2}$,
$R_{h,\sigma\sigma}=
[\sin (q_{h,\sigma}a)/\sin (q_{e,\sigma}a)]\vert
r_{h,\sigma\sigma} \vert^{2}$,
$R_{e,\sigma\sigma}=\vert r_{e,\sigma\sigma}\vert^{2}$,
and $R_{e,\underline{\sigma}\sigma}
=[\sin (q_{e,\bar{\sigma}}a)/\sin (q_{e,\sigma}a)]
\vert r_{e,\underline{\sigma}\sigma}\vert^{2}$
with $q$'s the wave vectors associated with different types of
electrons and holes.
In Fig.~\ref{FIG:FP-K}, we plot the conductance spectrum
and the reflection coefficients for various values of the spin-flip
interfacial scattering strength $U_{1}$ in a ND$_{xy}$ junction with the
spin-nonflip interfacial scattering strength $U_0=0.625D$. The transverse 
momentum is taken to be $k_y=\pi/3\sqrt{2}a$.
In the absence of the spin-flip scattering, i.e., $U_1=0$, the
coefficients $R_{e,\underline{\sigma}\sigma}$ and $R_{h,\sigma\sigma}$
vanish, and $R_{h,\underline{\sigma}\sigma}$ decreases while
$R_{e,\sigma\sigma}$ increases monotonically with the bias within the
effective energy gap $\vert \Delta_{\bf k}\vert =2\Delta_{d0}$.
Especially, $R_{h,\underline{\sigma}\sigma}=1$ and
$R_{e,\sigma\sigma}=0$ at $E=0$, which accounts for the appearance of the
ZBCP. (see Fig.~\ref{FIG:FP-K}(a))
The effects of the spin-flip scattering on the conductance spectrum   
depends on its strength in detail. For a relative small
value of the spin-flip scattering ($U_1=0.125D$),
$R_{e,\underline{\sigma}\sigma}$ and $R_{h,\sigma\sigma}$
decreases monotonically with the bias. $R_{e,\sigma\sigma}$ is finite
and $R_{h,\underline{\sigma}\sigma}$ is depressed at $E=0$, which leads to a
suppressed ZBCP. (see Fig.~\ref{FIG:FP-K}(b))
As the spin-flip scattering strength is further increased ($U_1=0.25D$),
$R_{h,\underline{\sigma}\sigma}$ varies nonmonotonically, it first 
increases  
to reach a maximum and then decreases with the bias. In addition, the
zero-bias $R_{e,\underline{\sigma}\sigma}$ and $R_{e,\sigma\sigma}$
are enhanced. Consequently, a  flat conductance maximum at a finite
bias shows up. (see Fig.~\ref{FIG:FP-K})(c)) As the
spin-flip scattering strength is comparable to the spin-nonflip  
part, the complementary behavior in the variation
between $R_{e,\underline{\sigma}\sigma}$ and $R_{e,\sigma\sigma}$
and weak bias-dependence of $R_{h,\underline{\sigma}\sigma}$ (highly 
suppressed
within the gap) and $R_{h,\sigma\sigma}$ leads to an almost constant
conductance spectrum. (see Fig.~\ref{FIG:FP-K}(d)) If the
spin-flip scattering is much stronger than
the spin-nonflip part ($U_1=1D$), the induction of a peak at finite
bias in both $R_{h,\underline{\sigma}\sigma}$ and $R_{h,\sigma\sigma}$
causes a finite-bias conductance peak. (see Fig.~\ref{FIG:FP-K}(e))
Whether such a extremely strong spin-flip scattering compared with 
the spin-nonflip scattering exits experimentally is unclear 
and we will not discuss this extreme limit further.
As shown in Fig.~\ref{FIG:FPND-AV},
the ZBCP can be completely
depressed in the averaged conductance spectrum 
of a ND$_{xy}$ junction ($U_0=0.625D$) 
by the spin-flip scattering.  
Figure~\ref{FIG:FPFD-AV} plots the averaged conductance
for various values of $U_1$ in a FD$_{xy}$ junction with $U_0=0.625D$ 
and $h_0=0.475D$. Since the spin-flip interfacial scattering tends to 
spoil the pre-oriented spin direction  of conduction electrons incident from 
the ferromagnet, it seriously influences the conductance spectrum. 
In particular, the splitting of the ZBCP induced by the exchange field 
is washed out in the presence of a strong spin-flip interfacial 
scattering so that the conductance spectrum
becomes completely structureless.
In addition, as shown in Figure~\ref{FIG:FPFD-AV}, 
the suppression of the conductance at the region 
away 
from zero bias by the exchange field is reduced by a strong spin-flip 
interfacial scattering.

\section{Local magnetic moment}
\label{SEC:LMM}
It has been predicted~\cite{Hu94} that besides the ZBCP in the 
quasiparticle tunneling, one of the other consequences of midgap states 
is the possibility of a sizable magnetic moment at the $\{110\}$ surface 
of the $d_{x_{a}^{2}-x_{b}^{2}}$-wave superconductor. 
Actually, the splitting of the ZBCP in the FD$_{xy}$ junction with a 
strong exchange field or in the ND$_{xy}$ junction with an 
in-plane magnetic field has supported this prediction. In a recent 
theoretical work,~\cite{HY98} a 
formal expression for the magnetic moment has been given, 
but a serious calculation of this quantity has not been done.
In this section, we give a detailed analysis of the local magnetic 
moment (LMM). 

The average 
electron density for each spin direction is given by \begin{equation}
\langle n_{{\bf i}\sigma} \rangle =\sum_{\nu} 
\{ 
\vert u_{{\bf i}\sigma}^{\nu}\vert^{2}f(E_{\nu})
+\vert v_{{\bf i}\sigma}^{\nu}\vert^{2}[1-f(E_{\nu})] 
\}\;,
\end{equation}
where $f(E)=[1+\exp (E/T)]^{-1}$ is the Fermi distribution function.
The local magnetic moment can be defined as 
\begin{equation}
m=-\mu_{B} (n_{{\bf i}\uparrow}-n_{{\bf i}\downarrow})\;.
\end{equation}
For the FD$_{xy}$ junction at temperature $T=0.08\Delta_{d0}$, 
the LMM at the distance $x=a/\sqrt{2}$ away from the 
interface in the superconducting region is found to be:
When  $h_{0}=0.5D$, 
$m(x=a/\sqrt{2})=0.074\mu_{B}$ for $U_{0}=0.625D$, and 
$0.008\mu_{B}$ for $U_{0}=2.5D$.
Corresponding to the splitting of the ZBCP in the FD$_{xy}$ junction, 
the  exchange-field induced local magnetization 
at the surface of $d$-wave superconductor is sensitive to 
the interfacial barrier strength. 
In Fig.~\ref{FIG:GMMa}, we plot the magnetic-field dependence of the LMM 
at $x=a/\sqrt{2}$ for different temperatures. At low fields, the LMM 
varies linearly with the field. The slope increases with the decreased 
temperature. At higher fields (about six times of the temperature) so that 
the width of 
the midgap peak in density of states is surpassed, the LMM begins to 
saturate. 
In contrast, for the FS junction or the normal-metal--$s$-wave 
superconductor junction with a Zeeman coupling, we find that 
the LMM is almost zero. 
In Fig.~\ref{FIG:GMMb}, we plot the spatial variation of the LMM into the 
superconducting region of the ND$_{xy}$ junction with $U_{0}=2.5D$ and 
$\mu_{B}B=0.4\Delta_{d0}$. The temperature $T=0.08\Delta_{d0}$. It is 
shown that the LMM has nonzero values only at the sublattice 
$x=(2n+1)a/\sqrt{2}$ with $n$ being non-negative integer, and it decays 
into the bulk of the superconductor. This interesting behavior directly 
reflects the existence of zero-energy peak at $x=(2n+1)a/\sqrt{2}$ 
and the absence at $x=2na/\sqrt{2}$ as the chemical potential 
$\mu=0$.~\cite{TTYK98} 
This feature is special to the lattice model.

\section{Conclusions}
\label{SEC:SUMMARY}
In conclusion, we have presented a unified theory for the proximity effect, 
quasiparticle transport, and local magnetic moment in the FD$_{xy}$ 
junctions by solving the Bogoliubov-de Gennes 
equations within an extended Hubbard 
model. As a comparison, the calculations are also made for the FS 
junctions. The energy gap appearing in the BdG equations have been 
determined self-consistently by using exact diagonalization technique. 
It is found that the proximity induced 
order parameter oscillates in the ferromagnetic region but is almost 
unchanged  in the superconducting region   
by the exchange field. The modulation period of the proximity induced order 
parameter is shortened by the exchange field but the oscillation 
amplitude is decreased with the interfacial scattering. 
Once the superconducting energy gap for various interfacial scattering 
potentials is determined
self-consistently, a transfer matrix method has been proposed to 
calculate the the subgap conductance within a scattering 
approach. We find that the subgap conductance is suppressed by the
spin splitting of the energy band in the ferromagnet.
For a ballistic FD$_{xy}$ junction, a conductance dip is exhibited with 
strong exchange fields. In the presence of interfacial scattering, the ZBCP 
is split by the strong exchange field. The degree of this splitting 
depends on the barrier strength. In contrast, the ZBCP can be split very 
easily by an in-plane magnetic field  due to the local nature of the 
Zeeman coupling and the range of splitting is independent of the barrier 
strength. In addition, we also show that the spin-flip interfacial 
scattering can seriously influence the quasiparticle transport 
properties. As one of the consequences of the midgap states, a sizable 
local magnetic moment in the FD$_{xy}$ junction or in the ND$_{xy}$ 
junction in the presence of the Zeeman coupling has been found. 
Inspired by the observation of the zero-bias conductance dip 
in the ballistic ferromagnet--high-$T_{c}$ superconductor 
junctions, we believe that the other interesting behaviors predicted in 
this paper are also experimentally accessible.  

\acknowledgments
This work was supported by the Texas Center for 
Superconductivity at the University of Houston, 
by the Robert A. Welch Foundation, 
and by the grant NSF-INT-9724809. 

\appendix
\section{Symmetry property of the conductance spectrum}
\label{SEC:APPD}
The eigenstates of the BdG equations given by Eq.~(\ref{EQ:BdG}) exist in 
pairs: Each eigenstate from Eq.~(\ref{EQ:BdGa}) is related to a counterpart  
from Eq.~(\ref{EQ:BdGb}) as: 
\begin{equation}
\left( 
\begin{array}{c}
u_{\downarrow} \\
v_{\uparrow}
\end{array}
\right)_{-E}=
\left( 
\begin{array}{c}
-v_{\downarrow}^{*} \\
u_{\uparrow}^{*}
\end{array}
\right)_{E}\;.
\end{equation}
This mirror image property makes the conductance spectrum is symmetric about 
the zero of bias. Suppose we have a incident, and outgoing waves as  
solutions to  Eq.~(\ref{EQ:BdGa}) with energy $E$
$$
\left( \begin{array}{c}
a_{in}e^{iq_{e\uparrow}x} \\
b_{in}e^{-iq_{h\downarrow}x}
\end{array}
\right)
$$ 
and
$$
\left( \begin{array}{c}
a_{out}e^{-iq_{e\uparrow}x} \\
b_{out}e^{iq_{h\downarrow}x}
\end{array}
\right)\;.
$$  
The amplitudes of the incoming and outgoing waves are then connected by 
the scattering matrix:
\begin{equation}
\left( \begin{array}{cc}
r_{e\uparrow,e\uparrow} & r_{e\uparrow,h\downarrow}\\
r_{h\downarrow,e\uparrow} & r_{h\downarrow,h\downarrow} 
\end{array}
\right)_{E} 
\left( \begin{array}{c}
a_{in} \\ b_{in} \end{array} \right)
=\left( \begin{array}{c} 
a_{out} \\ b_{out}
\end{array}
\right) \;.
\label{EQ:S-A}
\end{equation}
From the mirror image property of the eigenfunctions, the incident and 
outgoing waves at energy $-E$ must be given by 
$$
\left( \begin{array}{c}
-b_{in}^{*}e^{iq_{h\downarrow}x}\\
a_{in}^{*}e^{-iq_{e\uparrow}x} 
\end{array}
\right)
$$ 
and 
$$
\left( \begin{array}{c}
-b_{out}^{*}e^{-iq_{h\downarrow}x} \\ 
a_{out}^{*}e^{iq_{e\uparrow}x} 
\end{array}
\right)\;.
$$ 
These wavefunctions is related by the scattering matrix at energy $-E$:
\begin{equation}
\left( \begin{array}{cc}
r_{e\downarrow,e\downarrow} & r_{e\downarrow,h\uparrow}\\
r_{h\uparrow,e\downarrow} & r_{h\uparrow,h\uparrow} 
\end{array}
\right)_{-E} 
\left( \begin{array}{c}
-b_{in}^{*} \\ a_{in}^{*} \end{array} \right)
=\left( \begin{array}{c} 
-b_{out}^{*} \\ a_{out}^{*}
\end{array}
\right) \;.
\end{equation} 
The above equation can be rewritten as:
\begin{equation}
\left( \begin{array}{cc}
r_{h\uparrow,h\uparrow} & -r_{h\uparrow,e\downarrow}\\
 -r_{e\downarrow,h\uparrow}& r_{e\downarrow,e\downarrow} 
\end{array}
\right)_{-E} 
\left( \begin{array}{c}
a_{in}^{*} \\ b_{in}^{*} \end{array} \right)
=\left( \begin{array}{c} 
a_{out}^{*} \\ b_{out}^{*}
\end{array}
\right) \;.
\label{EQ:S-B}
\end{equation}  
Comparing Eq.~\ref{EQ:S-B} with Eq.~\ref{EQ:S-A}, we find the relation:
\begin{equation}
\begin{array}{l}
r_{e\uparrow,e\uparrow}(E)= [r_{h\uparrow,h\uparrow}(-E)]^{*} \;, \\
r_{e\uparrow,h\downarrow}(E)=[-r_{h\uparrow,e\downarrow}(-E)]^{*}\;, \\
r_{h\downarrow,e\uparrow}(E)= [-r_{e\downarrow,h\uparrow}(-E)]^{*}\;, \\
r_{h\downarrow,h\downarrow}(E)= [ r_{e\downarrow,e\downarrow}(-E)]^{*}\;. 
\end{array}
\end{equation} 
The symmetry properties of the reflection amplitudes yields the identity:
\begin{equation}
G_{\sigma}(E)=G_{\sigma}(-E)\;.  
\end{equation}
Therefore, we have extended the symmetry property~\cite{DBA96} to the 
magnetic case.

\begin{figure}
\caption{Schematic geometry of the ferromagnet--$s$-wave superconductor 
junction with a $\{100\}$-oriented interface (a) 
and the ferromagnet--$d$-wave superconductor junction with 
a $\{110\}$-oriented interface defined on a 
two-dimensional lattice. 
The filled circles represent the ionic 
positions in the 
ferromagnet and the filled squares represent the ionic positions in the 
superconductor. The interface layer is represented by empty circles.
${\bf x}_{a,b}$ are crystalline axes.
The propfiles of $s$-wave and $d$-wave order parameters are also shown, 
respectively.    
}
\label{FIG:FS}
\end{figure}

\begin{figure}
\caption{The spatial variation of order parameter 
for various values of exchange field in the FS 
junction without the interfacial scattering potential. 
Here $D=8t$ is the band width.}
\label{FIG:OP-SH}

\end{figure}

\begin{figure}
\caption{The spatial variation of order parameter for 
various values of exchange field in the FD$_{xy}$
junction without the interfacial scattering potential. }
\label{FIG:OP-DH}
\end{figure}

\begin{figure}
\caption{The spatial variation of order parameter 
for various values of interfacial scattering potential in the 
FS junction with the exchange field $h_{0}=0.125D$. 
}
\label{FIG:OP-SI}
\end{figure}

\begin{figure}
\caption{The spatial variation of order parameter 
for various values of interfacial scattering potential in the 
FD$_{xy}$ junction with $h_{0}=0.125D$. 
}
\label{FIG:OP-DI}
\end{figure}

\begin{figure}
\caption{
A schematic drawing of (a) the spin-split energy band in the 
ferromagnet within the continuum model and (b) the Andreev 
reflection process at the interface between the ferromagnet and 
superconductor: A beam of spin-up electrons incident with angle 
$\theta_{N}$ and energy within the gap are normally reflected as spin-up 
electrons and Andreev reflected as spin-down holes.  
The Andreev reflection angle $\theta_{A}$ is related to $\theta_{N}$ by 
the conservation of the momentum component parallel to the interface.
The thick solid line represents the interfacial scattering
layer. Also shown are the $d$-wave order parameter profile and the
angle $\alpha$ of the crystalline orientation with respect
to the interface. $E_{F}$ is the Fermi energy.
}
\label{FIG:AR}
\end{figure}

\begin{figure}
\caption{The differential conductance 
spectrum for various values of exchange field in the FS junction 
without the interfacial scattering potential.  }
\label{FIG:FS0} 
\end{figure}

\begin{figure}
\caption{The differential conductance 
spectrum for various values of exchange field
in the FD$_{xy}$ junction  
without the interfacial scattering potential.
 }
\label{FIG:FD0} 
\end{figure}

\begin{figure}
\caption{The differential conductance 
spectrum for various values of exchange field in the FS junction 
with $U_{0}=0.2D$.  
}
\label{FIG:FSI} 
\end{figure}

\begin{figure}
\caption{The differential conductance 
spectrum for various values of exchange field in the FD$_{xy}$ 
junction with $U_{0}=0.625D$.  
}
\label{FIG:FDI} 
\end{figure}

\begin{figure}
\caption{The differential conductance 
spectrum for various values of Zeeman coupling $\mu_{B}$ in the 
ND$_{xy}$ junction with $U_{0}=0.625D$.  
}
\label{FIG:FDB} 
\end{figure}

\begin{figure}  
\caption{   
The transverse-momentum-dependent differential conductance
spectrum and the reflection coefficients in a ND$_{xy}$ junction
for various values of spin-flip scattering strength $U_{1}=0$ (a),
$0.125D$ (b), $0.25D$ (c), $0.625D$ (d), and $1D$ (e).
The spin-nonflip scattering strength $U_0=0.625D$. The transverse momentum
$k_y=\pi/3\sqrt{2}a$. The effective energy gap for this momentum is $\vert
\Delta_{\bf k}\vert=2\Delta_{d0}$.
}
\label{FIG:FP-K}   
\end{figure}

\begin{figure}
\caption{
The averaged differential conductance
spectrum in a ND$_{xy}$ junction
for various values of spin-flip scattering strength.
$U_0=0.625D$.
}
\label{FIG:FPND-AV}
\end{figure}

\begin{figure}
\caption{
The averaged differential conductance
spectrum in a FD$_{xy}$ junction
for various values of spin-flip scattering strength.
$U_0=0.625D$ and $h_0=0.475D$.
}
\label{FIG:FPFD-AV}
\end{figure}

\begin{figure}
\caption{The magnetic-field dependence of the 
local magnetic moment at the distance $a/\sqrt{2}$ 
away from the interface in superconducting region of the ND$_{xy}$ 
junction at different temperatures. The interfacial scattering potential 
$U_{0}=2.5D$. }
\label{FIG:GMMa} 
\end{figure}

\begin{figure}
\caption{The spatial variation of the 
local magnetic moment into the superconducting region of the ND$_{xy}$ 
junction at $T=0.08\Delta_{d0}$ and $\mu_{B}B=0.4\Delta_{d0}$. 
The interfacial scattering potential 
$U_{0}=2.5D$. }
\label{FIG:GMMb} 
\end{figure}

\end{document}